\documentclass[12pt,epsfig]{iopart}
\usepackage{graphicx}
\usepackage{epsfig}

\begin{document}

\title{$\phi$ meson production in Au + Au collisions at
$\sqrt{s_{NN}}$ = 200 GeV}

\author{Debsankar Mukhopadhyay \it{for the PHENIX collaboration\footnote[1]{See Appendix for the full collaboration list.}}}

\address{  Department of Physics \& Astronomy, Vanderbilt University,
Nashville, TN 37235, USA
}

\begin{abstract}

We present the results on the mid-rapidity 
$\phi$ meson production in the $K^{+}K^{-}$ decay channel in 
Au + Au collisions at $\sqrt{s_{NN}}$ = 200 GeV measured by the PHENIX
experiment at RHIC. The spectral shape of the $\phi$ resonance, studied at 
different collision centralities, is consistent with the particle data book.
The transverse mass spectra are measured in four centrality bins.
The inverse slopes (T), yields (dN/dy) and particle ratios are studied as a function of centrality. The nuclear modification 
factor is measured through the ratio, $R_{CP}$, of central to peripheral 
yields normalized to the  number of  nucleon-nucleon  collisions. The  
 $R_{CP}$ of the $\phi$ mesons is less than unity and is comparable to that 
of pions rather than  $R_{CP} \sim 1$ observed for protons and anti-protons.  

\end{abstract}




\section{Introduction}

Heavy ion collisions at relativistic energies may lead to the formation of a  
Quark-Gluon Plasma (QGP) in which chiral symmetry is restored and nuclear 
matter is deconfined. Theory predicts{\cite{phi-theory}} that the chiral 
symmetry restoration may influence strongly the spectral shape 
(centroid and width) of the $\phi$ meson. Since the mass of the 
$\phi$ meson is close to twice the kaon mass, a modification in the 
$\phi$ mass centroid can lead to an observable imbalance in the 
branching ratios of the $\phi$ into kaon and di-lepton decay channels\cite{phikkee}. Consisting of $s\bar{s}$ pair, the $\phi$ production in heavy ion
collisions is sensitive to the strangeness enhancement.

The PHENIX experiment at RHIC has $\phi$ mass resolution
that is better than or comparable to the natural width of $\phi$. Thus, it 
can detect the predicted medium modifications of the  $\phi$ mass and width, 
should they occur.   The data presented here are from Au + Au collisions at 
$\sqrt{s_{NN}}$ = 200 GeV and are obtained using the $K^{+}K^{-}$ decay 
channel. We present the centrality dependence of line shape, yields, 
inverse slopes and particle ratios. The nuclear modification factor $R_{CP}$
 of $\phi$ mesons is also obtained and compared to that of pions and protons. 

\section{Experiment, data selection and techniques}

The present analysis is based on the measurements performed using
the east arm of the PHENIX  spectrometer {\cite{phnxtrk}}. Particles are 
identified by their mass after determining their momenta and velocities. The 
momentum measurements are provided by a multi-layer drift chamber and a 
layer of pad chambers. The velocities are obtained through 
time-of-flight (TOF) measurements. Two detector systems are used for TOF  
measurements: a TOF scintillator wall covering 
$\Delta\phi$ $\approx$ $\pi/8$ and a lead scintillator 
electromagnetic calorimeter ($\Delta\phi$ $\approx$ $\pi/4$). The 
PHENIX east arm covers pseudo-rapidity $|\Delta \eta| < 0.35$.
The collision vertex is measured by the Beam-beam counters (BBC) and the
centrality is determined by the BBC along with the Zero Degree Calorimeter. 
We have used 20 $\times$ $10^{6}$ minimum bias events within 
$|z_{vertex}|< 30 $ cm.

The $\phi$ mesons are reconstructed by forming 
the $K^{+}K^{-}$ invariant mass distributions. 
 A large combinatorial background is inherent to the same event $K^{+}K^{-}$ pair
invariant mass distribution. The combinatorial background is estimated by
the event mixing technique which is described in details at ref.{\cite{phirecdipali,phiprc}}. The $\phi$ signal is extracted by subtracting the 
combinatorial background from the same event pairs. 

\section{Results}

\subsection{Centroid and width analysis}

\begin{figure}[t]
\includegraphics[width=0.6\linewidth]{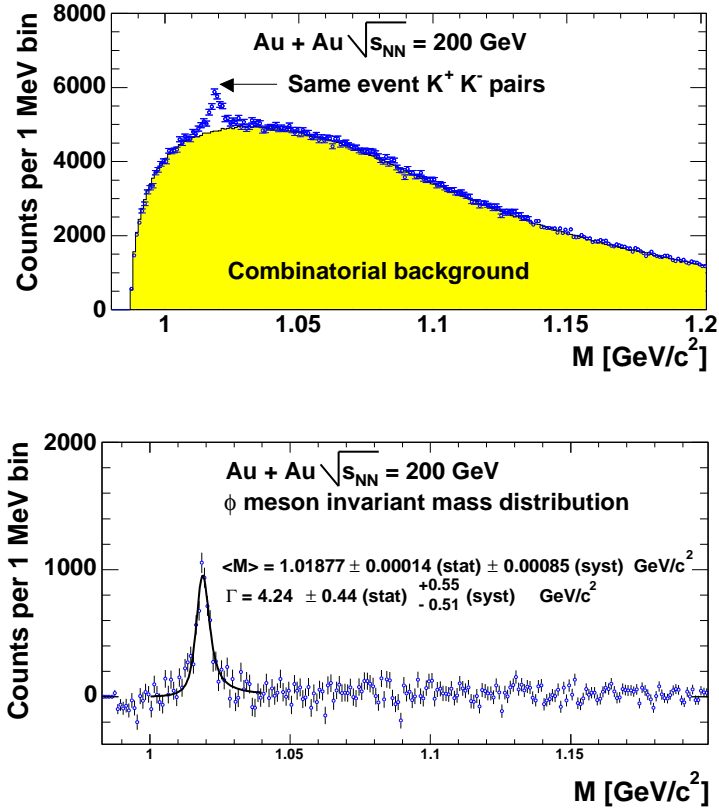}
\caption {\label{inv} Minimum-bias $\phi \rightarrow K^{+}K^{-}$
invariant mass spectrum using the kaons identified in the
PHENIX detector. The top panel shows the same event (circles) and combinatorial background $K^{+}K^{-}$ mass distributions. The bottom panel shows the
subtracted mass spectrum fitted with Relativistic Breit-Wigner function convolved
with the Gaussian experimental resolution function. }
\end{figure}

The Figure~\ref{inv} shows the $\phi$ meson invariant mass distribution
in $K^{+}K^{-}$ decay channel for the minimum bias
events. The upper panel 
of the figure shows the same event and combinatorial pair invariant mass 
distributions, while the lower panel presents the subtracted $\phi$ mass
spectrum where the combinatorial background beyond the $\phi$ peak is
essentially zero. The $\phi$ meson invariant mass 
peak is fitted with a relativistic Breit Wigner (RBW) distribution
convolved with a Gaussian experimental $\phi$ mass resolution function
with $\sigma$ = 1.0 MeV/$c^{2}$, which is determined by a Monte Carlo simulation. The fitted minimum bias mass
centroid and width, shown on the bottom panel of  the figure, 
are found to be consistent with the PDG values {\cite{PDG04}}.

\begin{figure*}[htb]
\includegraphics[width=0.7\linewidth]{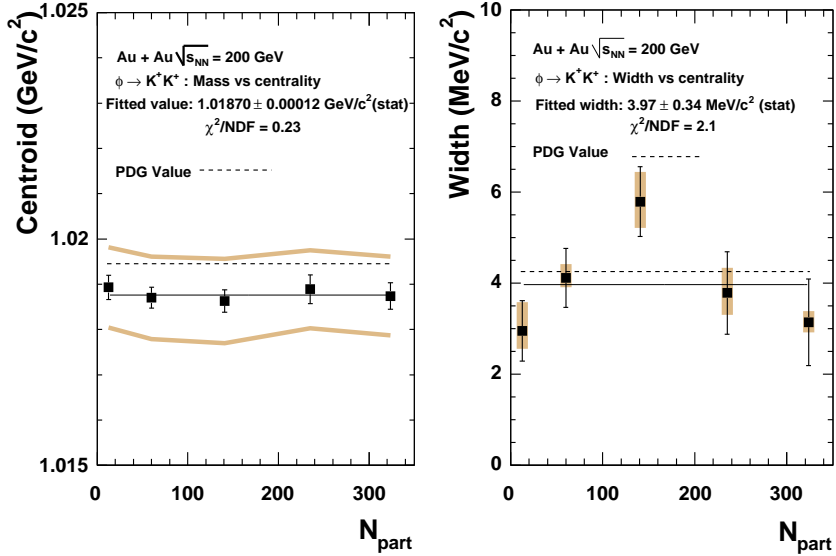}
\caption {\label{lineshape3} Centrality dependence of the $\phi$ mass
centroid (left) and $\phi$ intrinsic width (right).}

\end{figure*}

The centroids and widths are further studied as the function of centrality in 
Figure~\ref{lineshape3}.
The left panel of the figure
shows the centrality dependence of the fitted centroids. The upper and lower
1$\sigma$ systematic error limits are indicated. The dotted line shows the
PDG mass centroid. The solid line indicates the result obtained
with a one--parameter constant fit through the
measured data points. We observe that the centroid of the $\phi$ meson
resonance is consistent with the PDG values within errors.
The $\phi$ mass widths (right) are studied at different centralities.
The error bar on
each point shows the statistical error, while the bands on the points indicate
the systematic errors.  The dotted line shows the PDG $\phi$ mass
width. The solid line shows the results of the constant fit assumption to the
data points. Again, within the experimental errors, there
is no convincing evidence of a variation of the $\phi$~width as a function of centrality. 

\subsection{Transverse mass spectra, yields and ratios}

\begin{figure}[ht]
\includegraphics[width=0.4\linewidth]{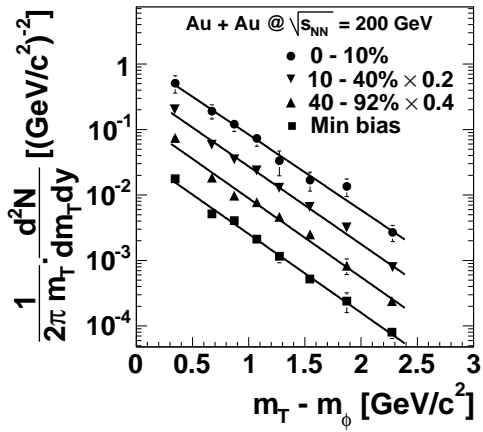}
 \caption{\label{mt2} $m_{T}$ spectra of $\phi$ mesons for 0 -- 10\%, 10 -- 40\%, 40 -- 92\% and minimum-bias 
(0 - 92\%) centrality classes.}
\end{figure}

The transverse mass spectra of the $\phi$ mesons are studied for the
minimum bias and three centrality classes, namely, 0 - 10\%, 10 - 40\% and 
40 - 92\%.  The invariant yields in different $m_{T}$ bins are obtained
by correcting the measured signal for geometrical acceptance and occupancy 
\cite{phiprc}. Each $m_{T}$ spectrum is then fitted with an exponential 
function in $m_{T}$ to extract dN/dy and inverse slope, T as two fitting 
parameters. Figure~\ref{mt2} shows the transverse mass spectra of $\phi$ 
mesons for above four centrality selections.

\begin{figure}[ht]
\includegraphics[width=0.7\linewidth]{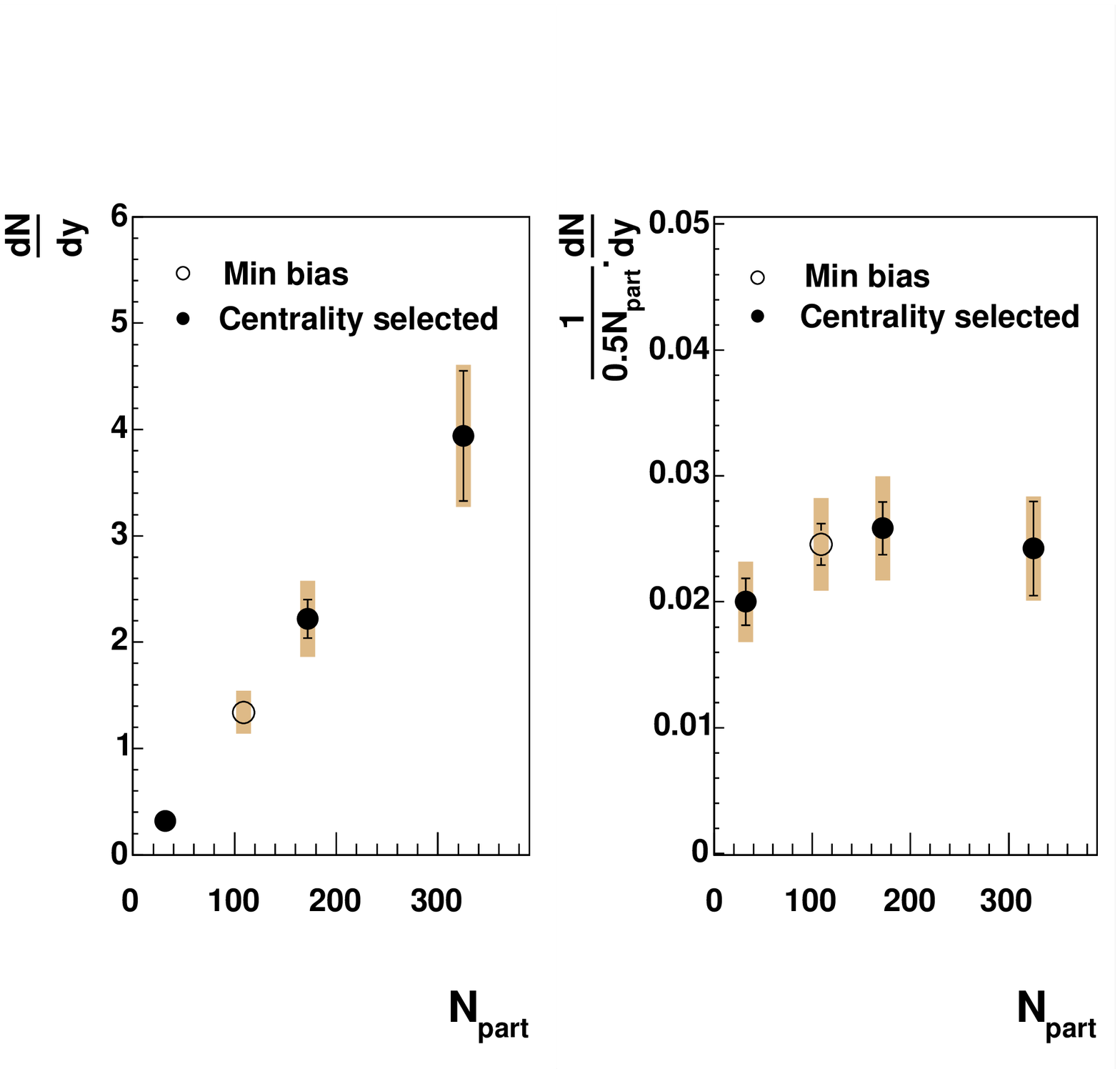}
   \caption{\label{cent1} Centrality dependence of $\phi$ yield at mid-rapidity. 
}
   \protect
\end{figure}

The centrality dependence of dN/dy is plotted in Figure~\ref{cent1}. The left
panel shows a steady increase in dN/dy with the number of participants. In the
right panel, the yield is normalized to the number of participant pairs to 
take into account the size of the system. Within the error bars this 
normalized rapidity density is approximately independent of centrality with a 
possible slight increase from peripheral to the mid-peripheral collisions. 
The trend is quite different from lower energy results measured at the
AGS. In~\cite{Back03}, the yield of $\phi$ was reported to be
increasing faster than linearly with the number of participants.

\begin{figure}[ht]
\includegraphics[width=0.7\linewidth]{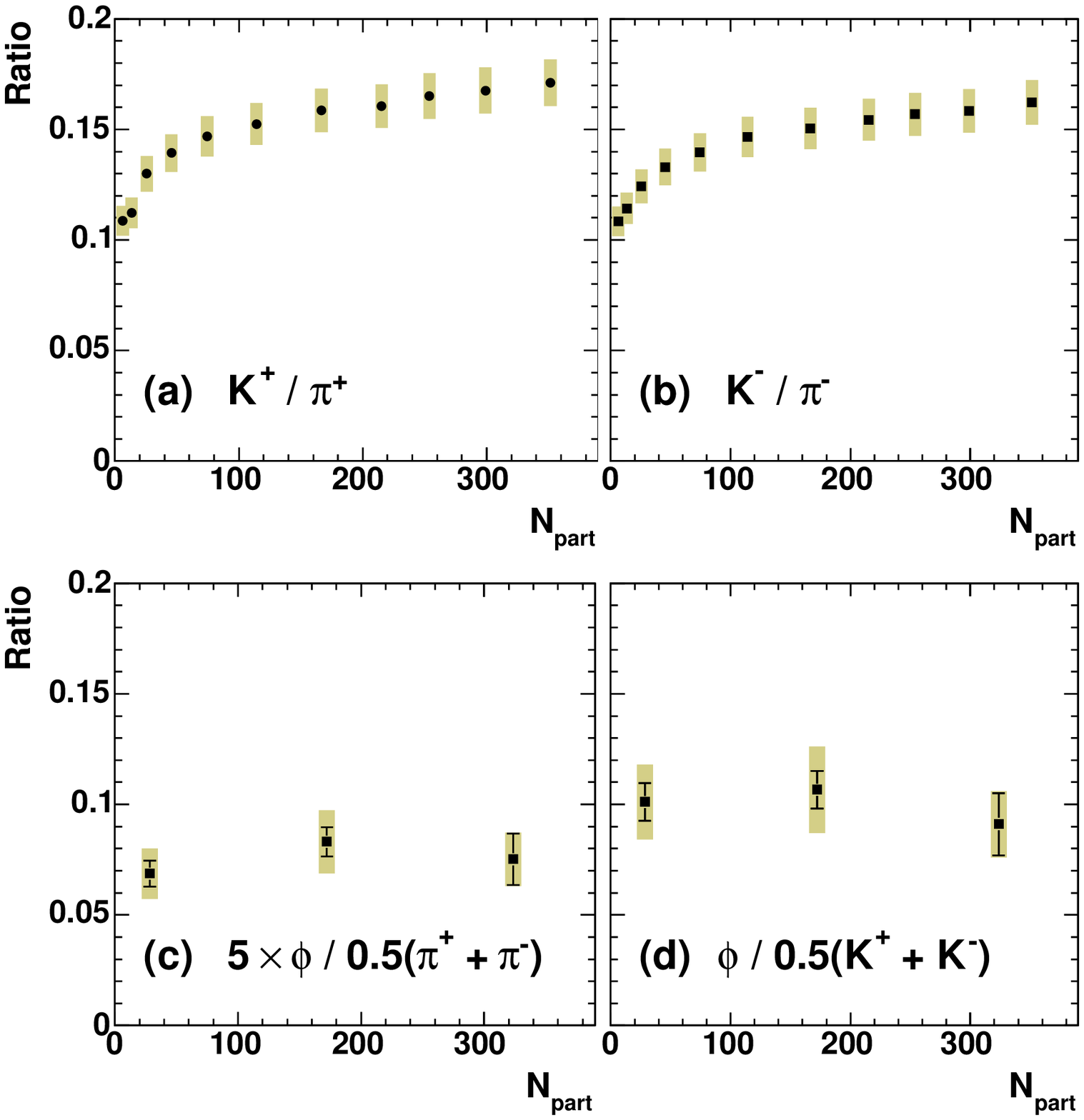}
\caption{\label{strange-npart} Centrality dependence of particle ratios
for (a) $K^{+} / \pi^{+}$, (b) $K^{-} / \pi^{-}$, (c) $\phi / 0.5
~(\pi^{+}+\pi^{-})$ (scaled by a factor of 5),
 (d)  $\phi / 0.5~(K^{+}+K^{-})$ in Au+Au collisions at $\sqrt{s_{NN}}$ = 200
GeV.}
\end{figure}

The extent and mechanism of strangeness enhancement are investigated via 
ratios	$\phi/\pi$ and $\phi/K$ at different centralities as shown in 
Figure~\ref{strange-npart} (c) - (d)
The $\phi / K$ ratio, in this limited number of centrality bins,
is approximately flat as a function of centrality. The possibility of 
structure in the $\phi/\pi$ ratio is difficult
to infer from our data within the error bars. For comparison, the ratios $K^{+} / \pi^{+} $ and  $K^{-} / \pi^{-}$ are also shown in Figure~\ref{strange-npart} (panels (a) and (b)). Since these ratios clearly increase with $N_{part}$,  $\phi / K$ and  $\phi / \pi $ can not be both flat.  Analysis of the 
significantly larger data set from Run4 of RHIC will be needed to map out
 the centrality dependence of these ratios.  

\begin{figure}[ht]
\includegraphics[width=0.5\linewidth]{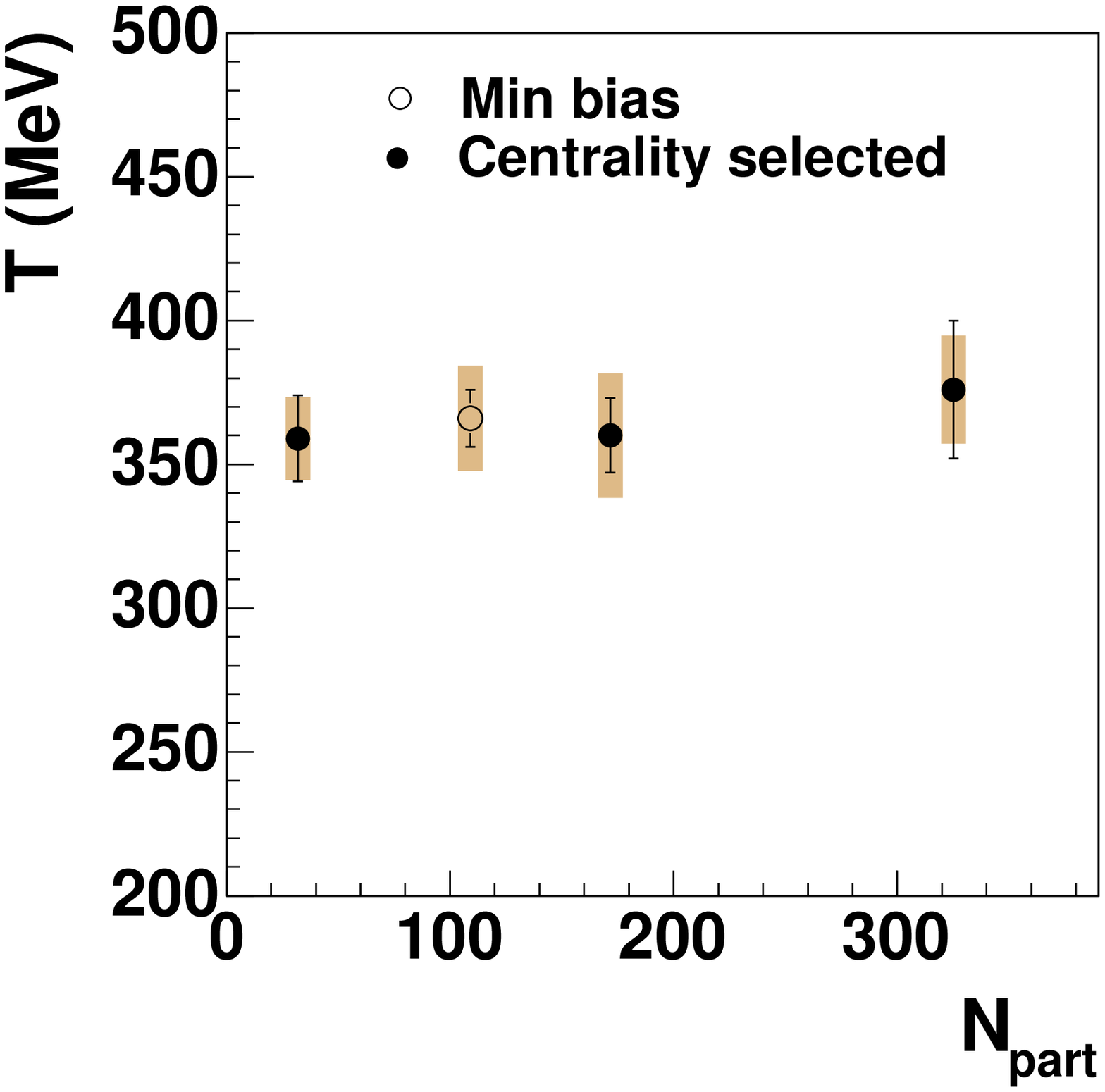}
   \caption{\label{cent2} Centrality dependence of the inverse slope, T.
}
   \protect
\end{figure}

The inverse slope, T, is found to be independent of collision centrality
as shown in Figure~\ref{cent2}. This result may be (wrongly) interpreted as 
an indication that the $\phi$-mesons do not participate in radial flow as 
the other hadrons. However, we note that our measurement does not extend 
down to low $m_{T}$ which is the region of the spectra that gets affected 
by radial flow. A hydrodynamics fit~\cite{phiprc} to the spectra of $\pi^{\pm}, K^{\pm}, p$ and $\overline{p}$ gives a satisfactory description of 
the $\phi$ spectra, too.    

\subsection{Nuclear Modification Factor, $R_{CP}$}
\begin{figure*}[ht]
\includegraphics[width=0.6\linewidth]{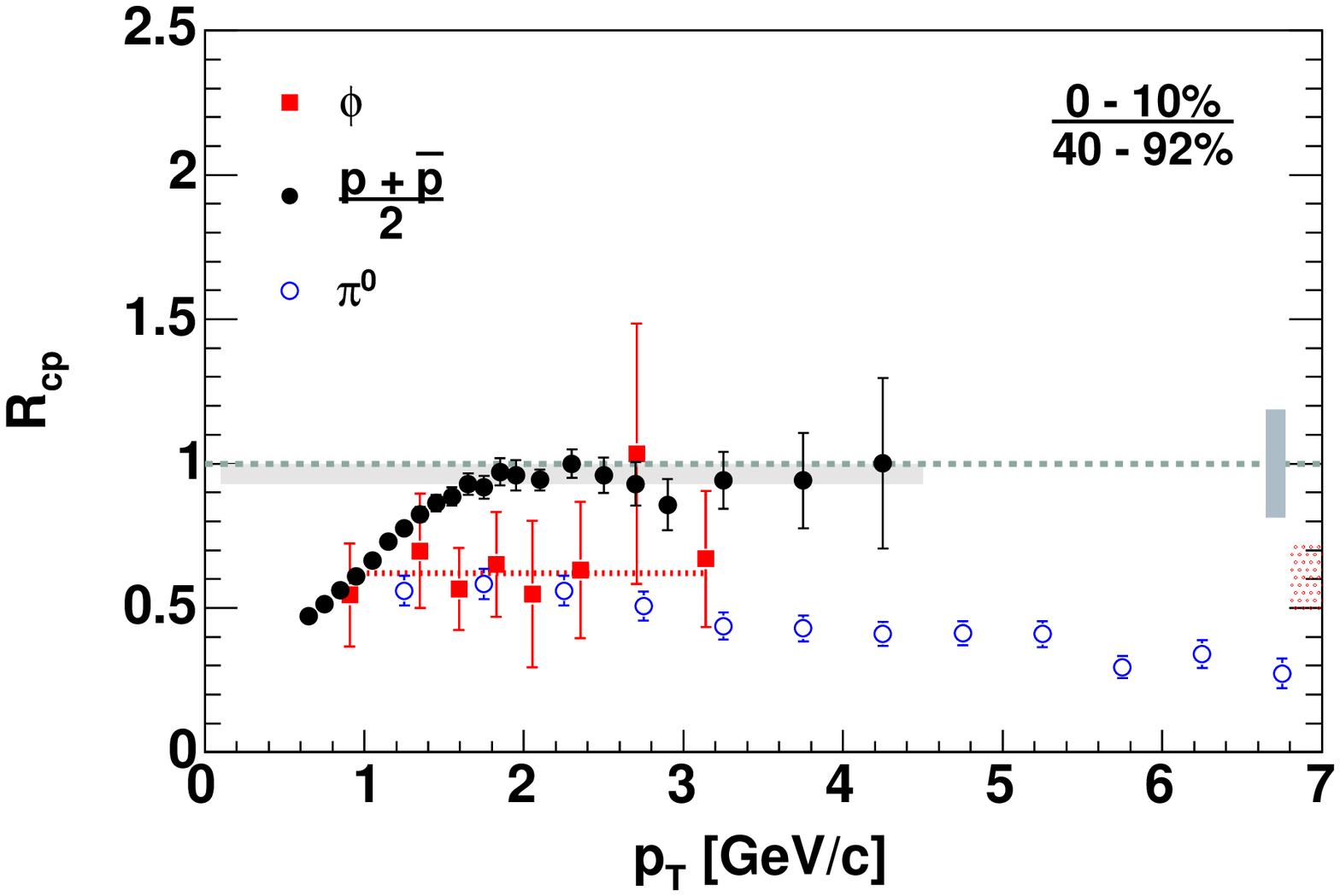}
\caption{\label{rcp} $N_{coll}$ scaled central to
   peripheral ratio $R_{CP}$ for $(p + \overline{p})$/2, $\pi^0$, and
   $\phi$. 
   The vertical dotted bar on the right represents the error on $N^{0-10\%}_{coll}
   /N^{40-92\%}_{coll}$ .  The shaded solid bar around $R_{CP} = 1$ represents
   12\% systematic error which can move the proton and/or $\phi$ points
   with respect to one another.  The dotted horizontal line at $R_{CP}$ = 0.62 is a
   straight line fit to the $\phi$ data.}
\protect
\end{figure*}

In order to understand the particle production mechanism at the
intermediate $p_{T}$ region, especially whether it is dominated
by the mass of the particles or their quark contents, it is 
important to study the nuclear modification factor, $R_{CP}$, which is
a ratio of the invariant yields normalized to the number
of collisions in central and peripheral collisions, as a function of $p_{T}$. 
The nuclear modification factor shows distinctive behavior for pions (mesons)
 and protons (baryons) {\cite{ppg015}}. Pions are suppressed, as expected 
from jet-quenching in the dense medium produced in the collisions, while 
(anti)proton production shows scaling typical for point-like processes in 
the absence of medium modifications. These seemingly contradictory results 
point to different dominant production mechanisms of pions and protons at 
moderately high $p_{T}$. The measurement of $\phi$ meson $R_{CP}$ provides 
a stringent test of theories that invoke the particle mass to explain pion/proton difference. In Figure~\ref{rcp}  we observe that $R_{CP}$ of the $\phi$ 
mesons is comparable to that of pions rather than protons, which have similar 
mass. This result may imply that the dominant hadron production mechanism at 
intermediate $p_{T}$ is sensitive to the quark content of the 
particles\cite{phiRcp,reco}.

\section{Summary}

We have performed a systematic study of the $\phi$ meson
production at mid-rapidity in Au+Au collisions at $\sqrt{s_{NN}}=200$~GeV.
The spectral shape of the $\phi$ resonance is found to be consistent with
PDG measured values. The yield, dN/dy, of the $\phi$ increases steadily with
centrality from 0.318$\pm$0.028(stat)$\pm$0.051(syst) in peripheral collisions
to 3.94$\pm$0.60(stat)$\pm$0.62(syst) in central collisions.
In the measured range ($m_{T}-m_{\phi} > 0.4$ GeV) , the inverse slope is 
essentially independent of the collision centrality. The nuclear modification factor for the $\phi$ mesons
is consistent with that of pions probably indicating a different
particle production process for mesons and baryons at intermediate $p_{T}$.


\section*{References}


\begin{thebibliography}{10}
\bibitem{phi-theory} K. Haglin, these proceedings; P. Koch, B. M\"uller and
J.~Rafelski, Phys. Rep. {\bf{142} 167 (1986}; 
\bibitem{phikkee} S. Pal, C. M. Ko and Z. Lin, Nucl. Phys. {\bf{A707}} 525 (2002).
\bibitem{phnxtrk} K. Adcox {\it et al.} (PHENIX Collaboration), Nucl. Instr.
Meth. {\bf{A499}} 489 (2003).
\bibitem{phirecdipali} D. Pal, these proceedings.
\bibitem{phiprc} PHENIX Collaboration, S.~S.~Adler {\it et al.}, nucl-ex/0410012

\bibitem{PDG04} S.~Eidelman {\it et al.},Phys. Lett. {\bf B592} 11 2004
\bibitem{Back03} B.~B.~Back {\it et al.} (E917 Collaboration), nucl-ex/0304017 (2003).
\bibitem{ppg015} PHENIX Collaboration, S.~S.~Adler {\it et al.}, Phys. Rev. Lett {\bf{91}} {172301} {2003}.).
\bibitem{phiRcp} J. Velkovska, J. Phys. {\bf{G 30}}, S835-S844 (2004), nucl-ex/0405013
\bibitem{reco} V. Greco, these proceedings; C. Nonaka, these proceedings; R. Fries, these proceedings; T. Chujo, these proceedings.
\end{thebibliography}
\end{document}